# Measuring eccentricity of binary black holes in GWTC-1 by using inspiral-only waveform

Shichao Wu, Zhoujian Cao⋆, Zong-Hong Zhu
*Department of Astronomy, Beijing Normal University, Beijing 100875, China*

**ABSTRACT**
In this paper, we estimate the eccentricity of the 10 BBHs in `GWTC-1` by using the inspiral-only BBH waveform template `EccentricFD`. Firstly, we test our method with simulated eccentric BBHs. Afterwards we apply the method to the real BBH gravitational wave data. We find that the BBHs in `GWTC-1` except GW151226, GW170608 and GW170729 admit very small eccentricity. Their upper limits on eccentricity range from 0.033 to 0.084 with 90% credible interval at the reference frequency 10 Hz. For GW151226, GW170608 and GW170729, the upper limits are higher than 0.1. The relatively large eccentricity of GW151226 and GW170729 is probably due to ignoring the $\chi_{\text{eff}}$ and low signal-to-noise ratio, and GW170608 is worthy of follow-up research. We also point out the limitations of the inspiral-only non-spinning waveform template in eccentricity measurement. The measurement of BBH eccentricity helps to understand its formation mechanism. With the increase in the number of BBH gravitational wave events and the more complete eccentric BBH waveform template, this will become a viable method in the near future.

## 1 INTRODUCTION

Since LIGO successfully detected GW150914 on September 14, 2015 (Abbott et al. 2016b,c,d), the door of gravitational wave astronomy opened. LIGO detected 3 gravitational wave events of binary black hole (BBH) coalescence in O1 (Abbott et al. 2016a). LIGO-Virgo Scientific Collaboration detected 8 new GW events in O2 (Abbott et al. 2019b), including 7 events of BBH and 1 event of binary neutron star (BNS). The first BNS event GW170817 (Abbott et al. 2017b) opened the door of multi-messenger astronomy (Abbott et al. 2017c).

The data of O1/O2 have been released (Abbott et al. 2019a). A catalog of O1/O2 GW events, `GWTC-1` has been reported (Abbott et al. 2019b). The PyCBC group reanalyzed the public data and verified the results of `GWTC-1`. Their results are included in the `1-OGC` catalog (Nitz et al. 2019c). After that, other groups also reanalyzed the public O1/O2 data (Antelis & Moreno 2019; Venumadhav et al. 2019; Stachie et al. 2020). Several new BBH candidates are reported (Zackay et al. 2019a,b; Gayathri et al. 2019; Venumadhav et al. 2019). Recently, PyCBC group has made some improvements on their PyCBC pipeline (Nitz et al. 2019a), including the correction for short time variations of detectors' noise power spectral density (PSD) and the instantaneous network sensitivity. Based on this improvement they partially validated some new BBH candidates of other groups. The result is reported in the `2-OGC` (Nitz et al. 2019a). No significant eccentric BNS candidate are found (Nitz et al. 2019b).

O3 observation has also begun in April 2019. Dozens of CBC candidates[1] have been found including GW190425 (Abbott et al. 2020) and GW190412 (The LIGO Scientific Collaboration & the Virgo Collaboration 2020).

The masses of BBHs detected by gravitational waves are significantly higher than that obtained by the observation of X-ray binaries (Kreidberg et al. 2012; Casares & Jonker 2014; Corral-Santana et al. 2016; Tetarenko et al. 2016; Abbott et al. 2016a). This fact raises a question about the origin of these black holes. It is difficult to answer how these binary black holes are formed. Current theories about the formation mechanism can be divided into two categories: *isolated binary evolution* and *dynamical formation*. With different formation mechanisms, the event rate, mass distribution, spin distribution, and eccentricity distribution of BBHs will differ. It's possible to use gravitational waves to determine these distributions and to better understand the formation mechanism.

An isolated binary is also called a "field binary". The binary system is already a binary star system at the stellar stage. In order to merge by radiating gravitational waves within the age of the Universe (Celoria et al. 2018), people have proposed several hypotheses: (1) the common envelope hypothesis (Livio & Soker 1988; Taam & Sandquist 2000; Dominik et al. 2012; Kruckow et al. 2016) assumes the two stars will go through a series of common envelope stages, and consequently the orbital distance will be shortened by the effect of friction force; (2) the chemically homogeneous hypothesis (Mandel & De Mink 2016; De Mink & Mandel 2016) can avoid the problems caused by the expansion of the predecessor star and will not annex another predecessor star due to the expansion; (3) the failed supernova may provide a fallback-driven mechanism to harden the binary (Tagawa et al. 2018).

---

⋆ corresponding author: zjcao@amt.ac.cn

[1] https://gracedb.ligo.org/superevents/public/O3/

The mechanism of dynamical formation is different from that of isolated binary. The component objects in the binary system are already compact objects before the formation of the binary system. In dense environments, such as young star clusters, globular clusters and galactic nuclei, compact objects may form a binary system due to the release of gravitational energy during the encounter. There are many works studied the dynamical formation mechanism (Sigurdsson & Hernquist 1993; Wen 2003; Antonini et al. 2016; O'Leary et al. 2016; Gondán et al. 2018b; Samsing 2018; Hoang et al. 2018; Fragione et al. 2019; Takátsy et al. 2019). For example the dynamical evolution of primordial black holes (BHs) in dense star clusters (O'Leary et al. 2006) and the scattering of stellar-mass black holes in galactic nuclei (O'Leary et al. 2009a) have been carefully studied.

The formation rate of BBH under different formation mechanisms will be different, and consequently will lead to differences in the expected BBH event rate observed through gravitational waves. LIGO-Virgo Scientific Collaboration analyzed 10 BBH events of GWTC-1, and the event rate is obtained by fitting the phenomenological model. They found the BBH merger rate is about $R = 53.2^{+55.8}_{-28.2}$ Gpc$^{-3}$yr$^{-1}$ (90% credible interval) (Abbott et al. 2019d).

The spin of the BBH will also be different under different formation mechanisms. Measuring the spin of the BBH by gravitational waves is also an effective method to distinguish the different formation mechanisms. The spin of individual compact objects in an isolated binary should be parallel to the orbital angular momentum of the binary system due to the common envelope. In the dynamical formation mechanism, the frequent interaction of compact objects in the dense environment makes the spin of the individual compact object randomly distributed. Since the spin of a single compact object is difficult to accurately measure by gravitational wave, the projection to the direction of orbital angular momentum, the effective spin $\chi_{\text{eff}}$, is usually measured (Abbott et al. 2019b). Many works (Vitale et al. 2015; Farr et al. 2017, 2018; Fernandez & Profumo 2019; Bouffanais et al. 2019; Roulet & Zaldarriaga 2019) have been paid to study how to use the distribution difference of $\chi_{\text{eff}}$ to distinguish different formation mechanisms.

The mass distribution of BBH formed under different formation mechanisms will also be different. According to the 10 BBHs in GWTC-1, LIGO-Virgo Scientific Collaboration obtains that the probability of the more massive black hole in the BBH that greater than 45 $M_\odot$ is not higher than 1%, and the power-law index of the mass spectrum is $\alpha = 1.3^{+1.4}_{-1.7}$ (Abbott et al. 2019d).

The eccentricity is also an important indicator to distinguish the formation mechanisms of BBHs. The orbit of an isolated BBH will be circularized by gravitational wave. When the gravitational wave signal enters the LIGO sensitive frequency band (10 Hz to 1000 Hz), the orbital eccentricity is expected to approach 0 (Peters 1964; Hinder et al. 2008; Kowalska et al. 2011). Differently the compact objects in a dense environment constantly disturb each other. When two black holes meet to form a binary black hole system, a large orbital eccentricity is expected. For example, they may meet in a parabolic orbit in the galactic nuclei, which results in a BBH with the eccentricity near 1 (O'Leary et al. 2009b). So measuring the eccentricity of BBH is an effective means to clearly distinguish the formation mechanisms.

Parameters estimation through gravitational wave detection requires accurate templates. In order to measure the eccentricity of BBH, accurate eccentric BBH gravitational wave templates are needed. The SEOBNRv3 (Pan et al. 2014; Taracchini et al. 2014; Babak et al. 2017) and IMRPhenomPv2 (Hannam et al. 2014; Khan et al. 2016; Husa et al. 2016) templates are extensively used by LIGO-Virgo Scientific Collaboration. But all these templates only describe circular BBH, consequently LIGO-Virgo Scientific Collaboration does not measure the eccentricity in the catalog GWTC-1 (Abbott et al. 2019b). LIGO-Virgo Scientific Collaboration recently used a non-template search pipeline cWB to search eccentric BBHs (Abbott et al. 2019e). Except the known BBHs in GWTC-1, no other BBH signal was found. Without template, cWB can not estimate the eccentricity of BBHs.

Currently there are many inspiral-only waveforms including x-model (Hinder et al. 2010), postcircular (PC) model (Yunes et al. 2009), EccentricFD (Huerta et al. 2014), TaylorF2e (Moore & Yunes 2019b). These templates assume small eccentricity and do not support spin. Huerta et al. (2018) developed ENIGMA model, which is a time domain, inspiral-merger-ringdown model that describes nonspinning BBHs that evolve on moderately eccentric orbits. Recently Cao & Han (2017) developed an eccentric BBH waveform SEOBNRE. This full inspiral-merger-ringdown waveform SEOBNRE also supports spins of black holes. A good match of SEOBNRE to the numerical relativity waveforms has been found in the range of eccentricity from 0 to 0.6 at reference frequency $Mf_0 = 0.002$ (Liu et al. 2020). But its generation speed is slow and cannot be directly used for the actual estimation of gravitational waves. This forces the authors Romero-Shaw et al. (2019) and Romero-Shaw et al. (2020) combined SEOBNRE and a reweight method to measure the eccentricity. As the authors pointed out, this method has some limitations. It requires the use of two gravitational wave templates, one without eccentricity (such as IMRPhenomD (Khan et al. 2016)) and one with eccentricity (such as SEOBNRE): firstly use the template without eccentricity for parameter estimation and get the posterior distribution without eccentricity, afterwards use the template with eccentricity to weight the sampling points obtained in the previous step to obtain the posterior distribution with eccentricity. The prerequisite of this method is that the two likelihood functions obtained by the two templates should be very similar (Payne et al. 2019). Otherwise the number of effective sampling points will be seriously reduced. The systematic error caused by using different templates needs more research (Abbott et al. 2017a; Ashton & Khan 2019).

Along with gravitational wave detection, parameter estimation methods can roughly be divided into two types. The first one is the Fisher information matrix (FIM) (Cutler & Flanagan 1994), which was widely used in the early days. The second one is the current mainstream, Markov chain Monte Carlo (MCMC) and nested sampling, which are stochastic sampling algorithms (Veitch et al. 2015). In the early articles about measuring eccentricity, the Fisher information matrix method was widely used (Sun et al. 2015; Ma et al. 2017; Nishizawa et al. 2017; Gondán et al. 2018a; Pan et al. 2019). Due to the limitations of FIM itself (Vallisneri 2008; Rodriguez et al. 2013), it needs the true value of parameters and high signal-to-noise ratio, so none of the above works used real gravitational wave data. In recent years, people began to use stochastic sampling algorithms such as MCMC and nested sampling for eccentricity measurement (Lower et al. 2018; Moore & Yunes 2019a). But these works still remain on simulated signals. Based on the EccentricFD waveform,

Lower et al. (2018) found the minimum measurable eccentricity of the HLV detector network at the design sensitivity is 0.05 at 10 Hz.

In this work, we present the first measurement of BBHs' eccentricity in GWTC-1 by using inspiral-only waveform combined with standard parameter estimation methods. Following (Lower et al. 2018) we use `EccentricFD` waveform. The eccentricity support of this waveform ranges from 0 to about 0.4 (Huerta et al. 2014). This waveform is also taken by the PyCBC group to search eccentric BNS (Nitz et al. 2019b). Based on the `EccentricFD` waveform, we find that the BBHs in `GWTC-1` except GW151226, GW170608 and GW170729 admit very small eccentricity. Their upper limits on eccentricity rang from 0.033 to 0.084 with 90% credible interval at the reference frequency 10 Hz. For GW151226, GW170608 and GW170729, the upper limits are higher than 0.1. The relatively large eccentricity of GW151226 and GW170729 is probably due to ignoring the $\chi_{\text{eff}}$ and low signal-to-noise ratio, and GW170608 is worthy of follow-up research.

At the same time we also find the limitations of using inspiral-only non-spinning waveform, mainly due to the low signal-to-noise ratio of matched filtering and ignoring the effect of $\chi_{\text{eff}}$. We verify this conclusion by simulating a large number of low signal-to-noise ratio BBH signals.

In this paper we use the geometric units ($G = c = 1$). The remainder of this paper is organized as follows. In the Section 2, we show the method used in this paper. We introduce `EccentricFD` and Bayesian parameter estimation used in this paper. Later we introduce the calculation method of the gravitational wave frequency of the last stable orbit $f_{\text{lso}}$ (in order to cut off the actual signal). In the Section 3, we test our method by simulating BBH signals. We use GW150914-like BBH signals for testing. We simulate 200 low signal-to-noise ratio BBH signals for Kolmogorov-Smirnov test to verify the accuracy of parameter estimation. We show the relationship between 90% credible interval and the signal-to-noise ratio of the detector network. In the Section 4, we show the measurement results of the `GWTC-1`. In the Section 5, we discuss the future development of eccentricity measurement method and its applications.

## 2 METHODS

In this section, we describe the method used in this paper. The Section 2.1 briefly review the `EccentricFD` waveform template. The Section 2.2 introduces our parameter estimation scheme. Since the `EccentricFD` template has only inspiral part, we need a cut-off frequency to drop the extra signal. The Section 2.3 explains our calculation method of the cut-off frequency.

### 2.1 EccentricFD waveform

`EccentricFD` is the name of a waveform template in LAL-Suite (LIGO Scientific Collaboration 2018), which corresponds to enhanced post-circular (EPC) model in the original paper (Huerta et al. 2014). The EPC model in the original paper is designed for a single detector. It calculates the strain of the single detector (in the detector frame). `EccentricFD` decomposes the two polarizations of the gravitational wave ($h_+$ and $h_\times$) and transforms from the detector frame to the geocentric frame. We denote the mass of the two compact objects as $m_{1,2}$, the total mass as $M = m_1 + m_2$, the symmetric mass ratio as $\eta = \frac{m_1 m_2}{M^2}$, and the chirp mass as $\mathcal{M} = \eta^{3/5} M$.

The EPC model is constructed by combining two approximations: the high-order quasi-circular post-Newtonian (PN) approximation and the leading-order post-circular approximation. The EPC model can be written as

$$h(f) = C \frac{\mathcal{M}^{5/6}}{D_L} f^{-7/6} \sum_{\ell=1}^{10} \xi_\ell \left(\frac{\ell}{2}\right)^{2/3} e^{-i(\pi/4 + \Psi_\ell)}, \quad (1)$$

$$\Psi_\ell = 2\pi f t_c - \ell \phi_c - \frac{\pi}{4} + \left(\frac{\ell}{2}\right)^{8/3} \frac{3}{128\eta (v_{\text{ecc}})^5} \sum_{i=0}^{i=7} a_i (v_{\text{ecc}})^i, \quad (2)$$

$$C = -\left(\frac{5}{384}\right)^{1/2} \pi^{-2/3}, \quad (3)$$

where $f$ is the frequency of GW, $D_L$ is the luminosity distance between the detector and the GW source, $\phi_c$ is the initial orbital phase, $t_c$ is the arrival time of GW singal in detector frame, and $v_{\text{ecc}}$ is the orbital velocity of the compact objects, which depends on $f$ and the initial eccentricity $e_0$. The explicit expression of $v_{\text{ecc}}$ is listed in Huerta et al. (2014) and the coefficients $a_i$ can be found in Eq. (3.18) of Buonanno et al. (2009). The amplitude $\xi_\ell$ depends on the two polar angles $\iota$ and $\beta$, the polarization angle $\psi$, the GW source location $\theta$ and $\phi$. The detail expression of $\xi_\ell$ can be found in Yunes et al. (2009).

The EPC model can be used until the last stable orbit (LSO) frequency

$$f_{\text{lso}} = \frac{1}{6\sqrt{6}\pi M}. \quad (4)$$

The EPC model will recover `TaylorF2` when the eccentricity is equal to zero (Huerta et al. 2014).

### 2.2 Gravitational wave parameter estimation

Bayesian method has been used for parameter estimation of gravitational waves (Finn & Chernoff 1993; Cutler & Flanagan 1994; Nicholson & Vecchio 1998; Christensen & Meyer 2001; Cornish & Crowder 2005), and Markov chain Monte Carlo (MCMC) method has become the mainstream method for gravitational wave parameter estimation (Van Der Sluys et al. 2008; Veitch & Vecchio 2010; Raymond 2012; Aasi et al. 2013; Veitch et al. 2015) in the past ten years. There are already some software packages dedicated to parameter estimation of gravitational waves, including `LALInference` (Veitch et al. 2015), `PyCBC Inference` (Biwer et al. 2019) and `Bilby` (Ashton et al. 2019). We use `Bilby` (Ashton et al. 2019) and `dynesty` sampler (Speagle 2019) in the current study.

After the search pipelines such as `cWB` (Klimenko & Mitselmakher 2004; Klimenko et al. 2005, 2008, 2016), `GstLAL` (Messick et al. 2017; Sachdev et al. 2019) and `PyCBC` (Nitz et al. 2018) have found a GW trigger, the corresponding data $\vec{d}(t)$ around that trigger time should be analyze more for the GW source parameters $\vec{\vartheta}$ under some model assumption $H$, say CBC model as considered in the current paper. According to the Bayes' theorem (Bayes 1763), we have

$$p(\vec{\vartheta}|\vec{d}(t), H) = \frac{p(\vec{d}(t)|\vec{\vartheta}, H) p(\vec{\vartheta}|H)}{p(\vec{d}(t)|H)}, \quad (5)$$

where $p(A|B)$ means the conditional probability of event $A$ under the condition of given event $B$, $p(\vec{\vartheta}|\vec{d}(t), H)$ is the posterior probability density of model parameters, $p(\vec{\vartheta}|H)$ is the prior probability density, $p(\vec{d}(t)|\vec{\vartheta}, H)$ is the likelihood function for the Gaussian noise (Wainstein & Zubakov 1970)

$$p(\vec{d}(t)|\vec{\vartheta}, H) = \exp\left[-\frac{1}{2}\sum_{i=1}^{N}\langle \tilde{n}_i(f)|\tilde{n}_i(f)\rangle\right] \quad (6)$$

$$= \exp\left[-\frac{1}{2}\sum_{i=1}^{N}\left\langle \tilde{d}_i(f) - \tilde{s}_i(f,\vec{\vartheta})|\tilde{d}_i(f) - \tilde{s}_i(f,\vec{\vartheta})\right\rangle\right], \quad (7)$$

where $N$ is the number of detectors in the detector network. The inner product is

$$\langle \tilde{a}_i(f)|\tilde{b}_i(f)\rangle = 4\Re \int_{f_{\min}}^{f_{\max}} \frac{\tilde{a}_i(f)\tilde{b}_i(f)}{S_n^{(i)}(f)}\mathrm{d}f, \quad (8)$$

where $S_n^{(i)}(f)$ is the power spectral density (PSD) of the $i$-th detector's noise. $\tilde{d}_i(f)$ and $\tilde{n}_i(f)$ are the frequency domain representations of the data and noise. $\tilde{s}_i(f,\vec{\vartheta})$ is the $i$-th detector's strain (GW waveform model's projection on $i$-th detector) in the frequency domain.

In order to get the posterior probability density of model parameters, we need some stochastic sampling algorithms, such as Markov chain Monte Carlo (MCMC) (Metropolis et al. 1953; Hastings 1970) and nested sampling (Skilling et al. 2006). In this paper, we use the nested sampler `dynesty` (Speagle 2019) for parameter estimation.

In this paper, we sample the chirp mass $\mathcal{M}$ and mass ratio $q$, instead of sampling the mass of a single black hole ($m_1,m_2$), because the gravitational wave waveform is mainly determined by the chirp mass, as we can see in the post-Newton approximation, so that the convergence efficiency would be higher. Other parameters are the eccentricity $e$, the luminosity distance $D_L$, the declination $\theta$ and the right ascension $\phi$, the inclination angle between the line of sight and the total angular momentum of BBH $\theta_{JN}$, the polarization angle $\psi$, the phase at coalescence $\phi_c$, and the coalescence time $t_c$. The range of the prior distribution of all parameters is shown in the Table 1. The prior used in this paper is based on the default BBH prior in `Bilby`, which covers the parameters of 10 BBHs in `GWTC-1`. Because `EccentricFD` doesn't support spin, we don't set the spin prior. The prior of eccentricity is based on the prior used by Lower et al. (2018) and Romero-Shaw et al. (2019). They used the log-uniform prior on eccentricity. When you do not know the order of magnitude of a parameter, the log-uniform prior is a good choice (see the bottom right corner of page two in Thrane & Talbot (2019)). Since we use the log-uniform prior, 0 is not included in the range. The lower limit used in this paper is lower than the prior used in Lower et al. (2018) and Romero-Shaw et al. (2019), and also lower than the minimum measurable eccentricity (at 10 Hz) of aLIGO at design sensitivity (Lower et al. 2018).

### 2.3 Cut-off frequency

When the waveform template coincides with the gravitational wave signal in the data, the likelihood function will obtain the maximum value, which means that the randomly sampling points will converge around the true parameters in the parameter space. The `EccentricFD` waveform template has only the inspiral part. In order to avoid the residual gravitational wave signal corresponding to the merger and ringdown part interfere the likelihood function, we need to set the cut-off frequency and remove the merger and ringdown part signal in frequency domain.

As mentioned in the Section 2.1, we use the gravitational wave frequency of the last stable orbit (LSO) as the cut-off frequency,

**Table 1.** The range of the prior distribution of all parameters. $t$ means the trigger time of BBH, we also set constrain on $m_1$, $m_1 \in [1.001398, 150]$. The eccentricity $e$ means the initial eccentricity at the reference frequency 10 Hz. "UniformSourceFrame" means uniform in comoving volume and source frame time.

| parameter | unit | prior | range | boundary |
|---|---|---|---|---|
| $\mathcal{M}$ | $M_\odot$ | Uniform | [2, 60] | reflective |
| $q$ | 1 | Uniform | [0.125, 1] | reflective |
| $e$ | 1 | LogUniform | [$10^{-10}$, 0.4] | reflective |
| $D_L$ | Mpc | UniformSourceFrame | [100, 5000] | reflective |
| $\phi$ | rad. | Uniform | [0, $2\pi$] | periodic |
| $\theta$ | rad. | Cosine | [$-\pi/2$, $\pi/2$] | reflective |
| $\theta_{JN}$ | rad. | Sine | [0, $\pi$] | reflective |
| $\psi$ | rad. | Uniform | [0, $\pi$] | periodic |
| $\phi_c$ | rad. | Uniform | [0, $2\pi$] | periodic |
| $t_c$ | s | Uniform | [$t$-0.1, $t$+0.1] | None |

$f_{\mathrm{lso}}$, which is also the cut-off frequency of the `EccentricFD` template. The $f_{\mathrm{lso}}$ depends on the total mass $M$ of the BBH. We use the public data from PyCBC group (Nitz et al. 2019a). The PyCBC group has done the parameter estimation on `2-OGC` by using `PyCBC Inference`. The `PyCBC Inference` uses parallel tempering sampler `ptemcee` (Vousden et al. 2016). We take the median values of $m_1$, $m_2$ (source frame) for the 10 BBHs in `2-OGC`, which correspond to the 10 BBHs of `GWTC-1`. Combined with the median values of redshift, we get the masses in geocentric frame, then we obtain the median value of total mass $M$. Based on this $M$ we calculate the median value of $f_{\mathrm{lso}}$ through (4). We list the results in the Table 2. In addition we also calculate the $f_{\mathrm{lso}}$ for the four new BBH candidates in `2-OGC` for later usage. In order to make sure that $f_{\mathrm{lso}}$ is safely away from the merger and ringdown part, we also plot the minimum, median and maximum value of $f_{\mathrm{lso}}$ in Figure 1 and Figure 2. We also show the frequency-domain detector strain of BBHs in `GWTC-1` reconstructed by taking the median value of `2-OGC` parameter estimation samples. Note that the waveform is generated by non-eccentric, precessing BBH template `IMRPhenomPv2` (Hannam et al. 2014; Khan et al. 2016; Husa et al. 2016), which is the inspiral-merger-ringdown model used by LIGO. Then we project the waveform onto L1 detector to get the frequency-domain detector strain. We also show the L1 detector's amplitude spectral density (ASD) used in our analysis. The ASD is caculated through the mean-median Welch method (Veitch et al. 2015; Chatziioannou et al. 2019) by taking the real detector noise near each GW event. We can see that even the maximum value is safely away from the merger and ringdown part of each GW event. The frequency-domain detector strain of H1 is very similar, only a slight difference in amplitude due to the difference in detectors' orientation.

We note that "inspiral-merger-ringdown (IMR) consistency test" (Ghosh et al. 2016, 2017) used another cut-off frequency $f_c$, which is also used to distinguish the inspiral part and the post-inspiral part of the signal. The $f_c$ corresponds to the innermost stable circular orbit GW frequency of the final Kerr black hole. Following the method of Healy et al. (2014), we calculate the median values of the mass and spin of the final Kerr black hole based on the public posterior data of `2-OGC`, we take the median values of $m_1$, $m_2$ (geocentric frame) for the 10 BBHs in `GWTC-1` as what we have done for $f_{\mathrm{lso}}$, then we caculate the median values of spin in the orbital angular momentum direction by using the median values of the dimensionless spin and the polar angle of the dimensionless spin, using the method of Healy et al. (2014), we caculate the median values of the mass and spin of the final Kerr black hole, then we calculate the $f_c$ based on the median values of the mass and spin of the final Kerr black hole. Our results are consistent with the results listed in the Table III of (Abbott et al. 2019c). For comparison we also list these $f_c$ in the Table 2.

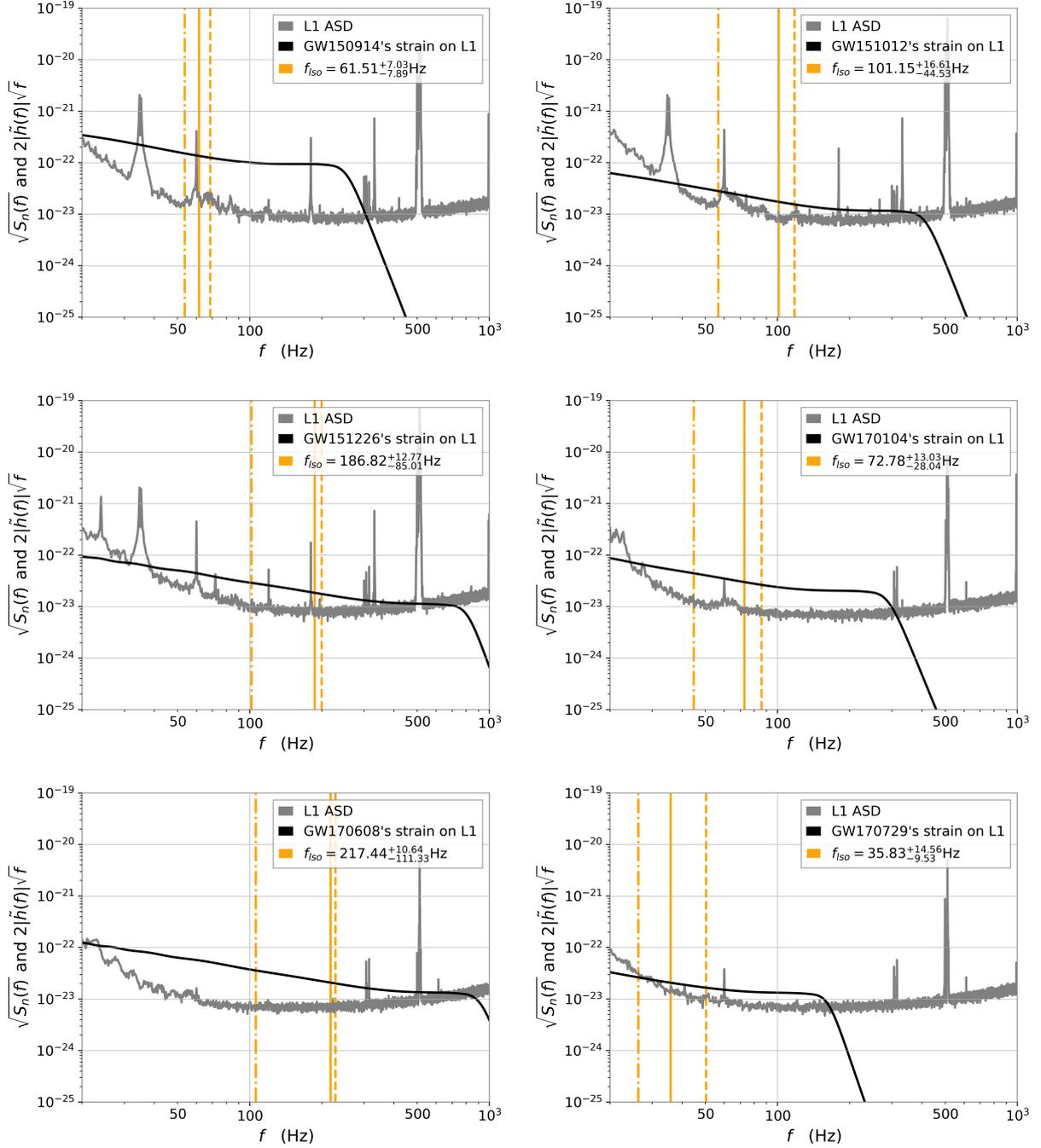

**Figure 1.** The minimum, median and maximum value of $f_{\text{Iso}}$ for GW150914, GW151012, GW151226, GW170104, GW170608, GW170729 in `GWTC-1`. The gray line is the L1 detector's amplitude spectral density (ASD) used in our analysis, the black line is the median frequency-domain detector strain of L1 generated by `IMRPhenomPv2`.

## 3 SIMULATED DATA STUDY

In order to verify the effectiveness of our method, we test our method against some simulated signals before performing parameter estimation on 10 real BBH gravitational wave signals in `GWTC-1`. In Section 3.1, we simulate two GW150914-like BBH signals. One admits eccentricity $e = 0.1$ and one without eccentricity ($e = 0$). We illustrate the effects of the eccentricity by plotting the time/frequency domain waveform. Then we do the parameter estimation of these two signals by using our method, and it finds that our method can accurately obtain the eccentricity. In order to statistically verify the validity of our method, we simulate 200 BBH events with eccentricity in Section 3.2. Our method passes the Kolmogorov-Smirnov test. Finally, we also discuss the relationship between the 90% credible interval of eccentricity and the signal-to-noise ratio of the detector network. The relationship indicates the limitation of inspiral-only waveform to measure eccentricity.

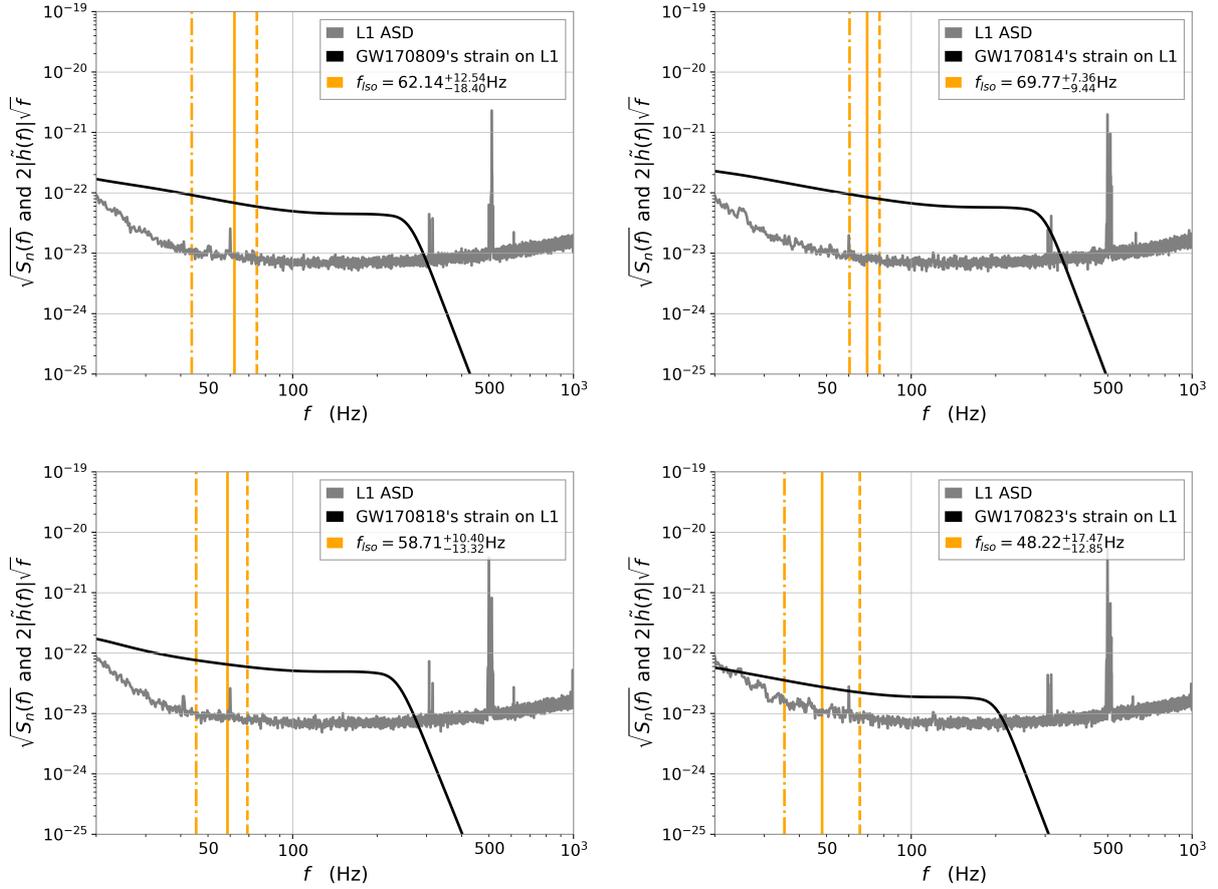

**Figure 2.** The minimum, median and maximum value of $f_{\rm iso}$ for GW170809, GW170814, GW170818, GW170823 in `GWTC-1`. The plot convention is the same as that of the Figure 1.

**Table 2.** The gravitational wave frequencies for the last stable orbit $f_{\rm iso}$ and the inner most circular orbit $f_c$. The results for the 10 BBH events and the 4 new candidates in `2-OGC`, which are marked with "*".

| event | $f_{\rm iso}$/Hz | $f_c$/Hz |
|---|---|---|
| GW150914 | 61.51 | 130.20 |
| GW151012 | 101.15 | 207.41 |
| GW151205* | 26.45 | 59.20 |
| GW151226 | 186.82 | 426.33 |
| GW170104 | 72.78 | 143.55 |
| GW170121* | 60.63 | 119.13 |
| GW170304* | 40.50 | 90.68 |
| GW170608 | 217.44 | 473.14 |
| GW170727* | 43.38 | 89.79 |
| GW170729 | 35.83 | 90.30 |
| GW170809 | 62.14 | 136.14 |
| GW170814 | 69.77 | 158.09 |
| GW170818 | 58.71 | 118.01 |
| GW170823 | 48.22 | 105.38 |

**Table 3.** The injected parameters of two GW150914-like BBHs. The eccentricity $e$ means the initial eccentricity at the reference frequency 10 Hz.

| parameter | unit | injected value |
|---|---|---|
| $m_1$ | $M_\odot$ | 35.0 |
| $m_2$ | $M_\odot$ | 30.0 |
| $e$ | 1 | 0.0 (0.1) |
| $D_L$ | Mpc | 440 |
| $\phi$ | rad. | 45 |
| $\theta$ | rad. | 5.73 |
| $\theta_{JN}$ | rad. | 0.4 |
| $\psi$ | rad. | 0.1 |
| $\phi_c$ | rad. | 1.2 |
| $t_c$ | s | 1126259462.422 |

### 3.1 GW150914-like BBHs test

We analyze two simulated GW150914-like BBH events in this subsection. The parameters of the two simulated gravitational wave events are shown in the Table 3. The values are set according to the Table I of Lower et al. (2018).

With the parameters listed in the Table 3, we use the complete eccentric waveform `SEOBNRE` (Cao & Han 2017; Liu et al. 2020) to generate the simulated signals for the purpose of illustration in the Figure 3, and we use `EccentricFD` in actual parameter estimation. We compare the time/frequency domain waveforms of these two simulated BBHs, as shown in Figure 3. In the top subplot, we can see that the time domain signal of eccentric BBH is shorter than the circular one. The eccentricity introduces amplitude modulation on the GW waveform. The difference is larger for the lower frequency part. In the bottom subplot, we can see the relative relationship between the strain caused by these two simulated BBHs and the sensitivity of the Advanced LIGO H1 detector (design sensitivity). We can also see the amplitude modulation in the low frequency part introduced by eccentricity. The three vertical lines represent 10 Hz, $f_{\rm Iso}$ and $f_{\rm c}$, respectively. According to the optimal SNR

$$\rho_{\rm opt}^2 = 4 \int_{f_{\rm min}}^{f_{\rm max}} \frac{|\tilde{h}(f)|^2}{S_n(f)} df = \int_{f_{\rm min}}^{f_{\rm max}} \frac{(2|\tilde{h}(f)|\sqrt{f})^2}{S_n(f)} d\ln(f), \quad (9)$$

the "area" between the GW strain and the ASD represents the optimal SNR. It is also the maximum SNR that matched filtering can achieve. According to the designed sensitivity of Advanced LIGO, we have $f_{\rm min} = 10$ Hz. Due to the cut-off frequency of the `EccentricFD` template used in this paper $f_{\rm max} = f_{\rm Iso}$, the matched filtering SNR will be smaller than the result of the complete eccentric BBH waveform. Differently the LIGO-Virgo Scientific Collaboration used $f_{\rm c}$ as the cut-off frequency in the inspiral-merger-ringdown (IMR) consistency test with the quasi-circular BBH templates. Therefore, the matched filtering SNR calculated in the Section 4 is different from the $\rho_{\rm insp}$ results in the Table III of Abbott et al. (2019c).

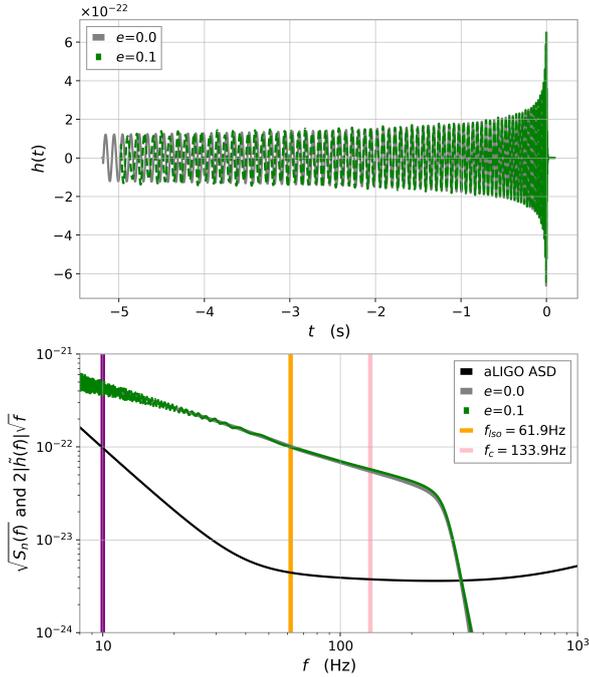

**Figure 3.** The time and frequency domain GW strain of the simulated GW150914-like BBHs for Advanced LIGO H1 detector.

We use the method described in the above section to calculate the posterior distribution of the source parameters. And the marginalized posterior distribution for the initial eccentricity at reference frequency 10 Hz can be got accordingly. We plot the results for the two simulated signals in the Figure 4. The vertical dash lines represent the true values. It can be seen that out method can accurately catch the eccentricity of BBHs. Other parameters can also be accurately got. We leave the posterior probability of the other primary parameters to the Appendix A.

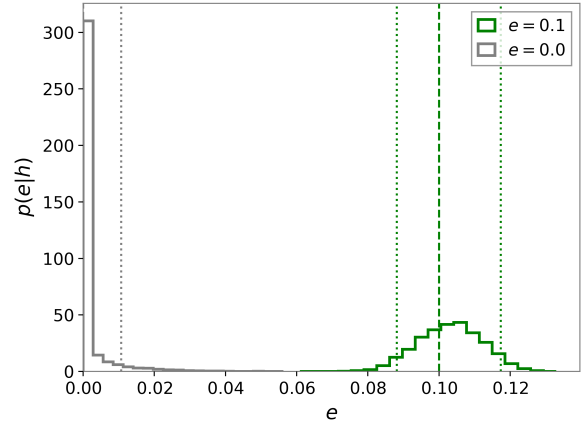

**Figure 4.** The 1D marginal posterior distribution of the eccentricity for the two simulated signals. The two dash lines represent the true values. The dotted lines represent the 90% credible intervals.

### 3.2 Kolmogorov-Smirnov (KS) test

We test the statistical property of our method in this subsection. According to the LIGO-Virgo Scientific Collaboration's standard, any gravitational wave parameter estimation method needs to pass the Kolmogorov-Smirnov (KS) test (Sidery et al. 2014; Veitch et al. 2015; Biwer et al. 2019; Thrane & Talbot 2019). For such KS test, we simulate 200 BBH events whose parameters are randomly drawn from the prior distribution listed in the Table 1. Here we use `EccentricFD` to simulate the signals. Then we inject these simulated signals into a detector network composed of the two Advanced LIGO detectors H1 and L1. Both detectors take their design sensitivity. Considering our matched filtering SNR on the real BBHs in `GWTC-1` will be relatively lower than the ones of Abbott et al. (2019c) because `EccentricFD` has only the inspiral part, we set the SNR range (6, 20) according to the $\rho_{\rm insp}$ in Table III of Abbott et al. (2019c). The final result is shown in the Figure 5. Our method can accurately recover the parameters of BBHs. The cumulative distribution curve of each parameter strictly follows the diagonal line and falls in the error range (the shadow area).

We have also investigated the relationship between the 90% credible interval and the signal-to-noise ratio of the detector network (Huang et al. 2018). The result is plotted in the Figure 6. The simulated data used here is the same as those in the Figure 5. Huang et al. (2018) fixed all parameters except the luminosity distance. It is different in the current work. We simulate signals with random parameters from the prior distributions. This is better to test the general capability of our method. The gray points represent the true eccentricity of 200 BBHs. The brown points represent the upper limit of the 90% credible interval calculated by our method. The brown line and the shaded area represent the fitting curve and the 90% confidence interval of the fitting, respectively. The black dash line represents the eccentricity's upper bound of the 90% interval of prior distribution, which is about 0.36. When the signal-to-noise ratio of the detector network increases, the 90% credible interval of the eccentricity becomes narrower. But within the range of the signal-to-noise ratio tested in this paper, the 90% credible interval of the eccentricity is generally around 0.1.

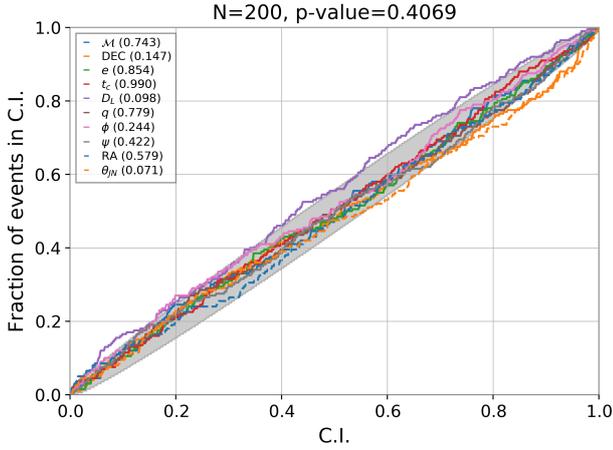

**Figure 5.** This plot shows the fraction of simulated BBH signals with parameter values within a credible interval as a function of credible interval. The diagonal line indicates the ideal 1-to-1 relation, which is expected if the parameter estimation method provides unbiased estimates of the BBH parameters. We perform the Kolmogorov-Smirnov (KS) test on each parameter (10 in total). The obtained two-tailed p-value indicate the consistency between the curves and the diagonal line. The combined p-value is showed in the plot title. The gray region indicates the fluctuations (90% credible interval) for the single parameter due to the finite sample size.

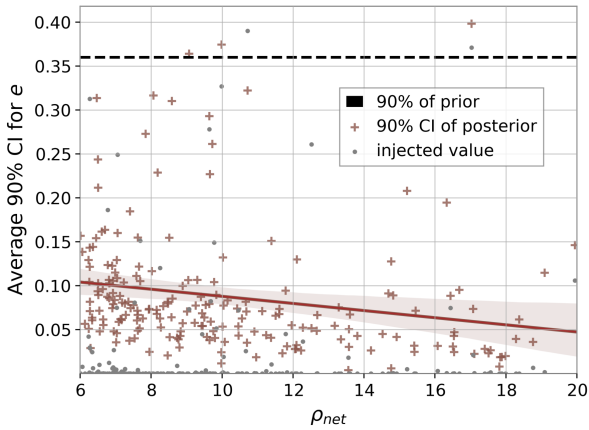

**Figure 6.** The relationship between the 90% credible interval and the signal-to-noise ratio of the detector network. The gray points represent the true eccentricity of 200 BBHs. The brown points represent the upper limit of the 90% credible interval calculated by our method. The brown line and the shaded area represent the fitting curve and the 90% confidence interval of the fitting, respectively. The black dash line represents the eccentricity's upper bound of the 90% interval of prior distribution, which is about 0.36.

**Table 4.** Measurement results for the 10 real BBH events in `GWTC-1`. $\rho_{net}$ is the median value of the network matched filtering signal-to-noise ratio calculated with `EccentricFD`. $e_{max}^{90}$ is the upper limit of the 90% credible interval.

| event | $\rho_{net}$ | $e_{max}^{90}$ |
| --- | --- | --- |
| GW150914 | 12.155 | 0.045 |
| GW151012 | 7.808 | 0.084 |
| GW151226 | 10.618 | 0.181 |
| GW170104 | 6.947 | 0.046 |
| GW170608 | 14.706 | 0.166 |
| GW170729 | 1.235 | 0.142 |
| GW170809 | 7.080 | 0.076 |
| GW170814 | 11.601 | 0.033 |
| GW170818 | 4.528 | 0.076 |
| GW170823 | 4.548 | 0.077 |

## 4 THE ECCENTRICITY MEASUREMENT OF BBHS IN GWTC-1

Now we apply our method to the real data of the 10 BBH events in `GWTC-1`. The sensitivity of Advanced LIGO and Advanced Virgo has not reached their respective design sensitivities during the O1 and O2 periods. There is a strong seismic noise below 20 Hz. So we follow the `GWTC-1` paper (Abbott et al. 2019b) and choose $f_{min} = 20$ Hz for the inner product. There is one exception. The H1 data of GW170608 has strong non-Gaussian noise below 30 Hz (Abbott et al. 2017d), so the $f_{min}$ of GW170608 in H1 data is set to 30 Hz. The $f_{Iso}$ is again chosen as the $f_{max}$ in the inner product. But note that we still use 10 Hz as the reference frequency for the initial eccentricity, which means we measure the eccentricity at 10 Hz of the GW. This is consistent with the reference frequency of the simulated signals in Section 3.2.

We plot the posterior distribution of eccentricity in the Figure 7 and Figure 8. The black line represents the upper limit of the 90% credible interval $e_{max}^{90}$. The gray histogram is drawn from the prior of eccentricity. We list the median value of the network matched filtering signal-to-noise ratio $\rho_{net}$ calculated with `EccentricFD` in the title after the GW event name. For clarity we also list $\rho_{net}$ and $e_{max}^{90}$ in the Table 4.

There are three BBHs, GW170729, GW170818, and GW170823, admitting a relatively low network signal-to-noise ratio $\rho_{net} < 6$. This is because the $f_{Iso}$ of these events is low. Their cut-off frequencies are respectively 35.83 Hz, 58.71 Hz and 48.22 Hz, closing to $f_{min} = 20$ Hz. The short integral frequency range results in a low signal-to-noise ratio.

There are three BBHs, GW151226, GW170608, and GW170729, admitting a relatively high upper bound of the eccentricity's 90% credible interval ($e_{max}^{90} > 0.1$). Next, we will discuss the possible reason.

Unlike `SEOBNRE`, the inspiral-only template `EccentricFD` does not consider the effect of spin. We need to consider whether the eccentricity measurement can be biased by the spin effect. This is related to the interpretation of the `GWTC-1` eccentricity measurement results. Here, we will focus on the impact of $\chi_{eff}$ on the eccentricity measurement. We use the quasi-circular ($e = 0$) precessing BBH template `IMRPhenomPv2` to simulate the injection signal, and then use `EccentricFD` to do parameter estimation. To see if it will be biased for the existence of spin effect. The priors of `IMRPhenomPv2` are listed in Table 5. The spin prior is based on `Bilby`'s BBH default prior, other priors are the same as Table 1. Because we only study the effect of $\chi_{eff}$ on the eccentricity measurement, we only simulate the case without precession. Combined with the $\chi_{\rm eff}$ measurement results of LIGO on `GWTC-1` (Abbott et al. 2019b), $\chi_{\rm eff}$ that can not be ignored (absolute value greater than 0.1) are all greater than zero. In order to make the results more statistically significant when computing resources are limited, we only simulate 100 BBH injections for the case in which $\chi_{\rm eff}$ is greater than 0. The signal-to-noise ratio range is the same as in Section 3. Then we measure the eccentricity of these injected signals by using this paper's method. The cut-off frequency $f_{\rm lso}$ is calculated from the injected parameters. The results are shown in Figure 9. The relationship between the width of the credible interval and the signal-to-noise ratio is inversely proportional (Cutler & Flanagan 1994), in order to reduce the impact of the different signal-to-noise ratio of the injected signal, we also plot the results of $e_{\rm max}^{90} \cdot \rho_{\rm net}$. If $\chi_{\rm eff}$ can't bias the eccentricity, $e_{\rm max}^{90}$ and $e_{\rm max}^{90} \cdot \rho_{\rm net}$ should be very small. Both figures clearly show that when $\chi_{\rm eff}$ is negligible (absolute value below 0.1), $\chi_{\rm eff}$ has very little effect on the eccentricity measurement. When $\chi_{\rm eff}$ is large, $\chi_{\rm eff}$ will have a strong impact on the eccentricity measurement. It is not a linear relationship, in some cases which $\chi_{\rm eff}$ is large, $e_{\rm max}^{90}$ is still small, the existence of $\chi_{\rm eff}$ may cause deviations in other parameters, this degeneracy problem needs to be further studied in the near future.

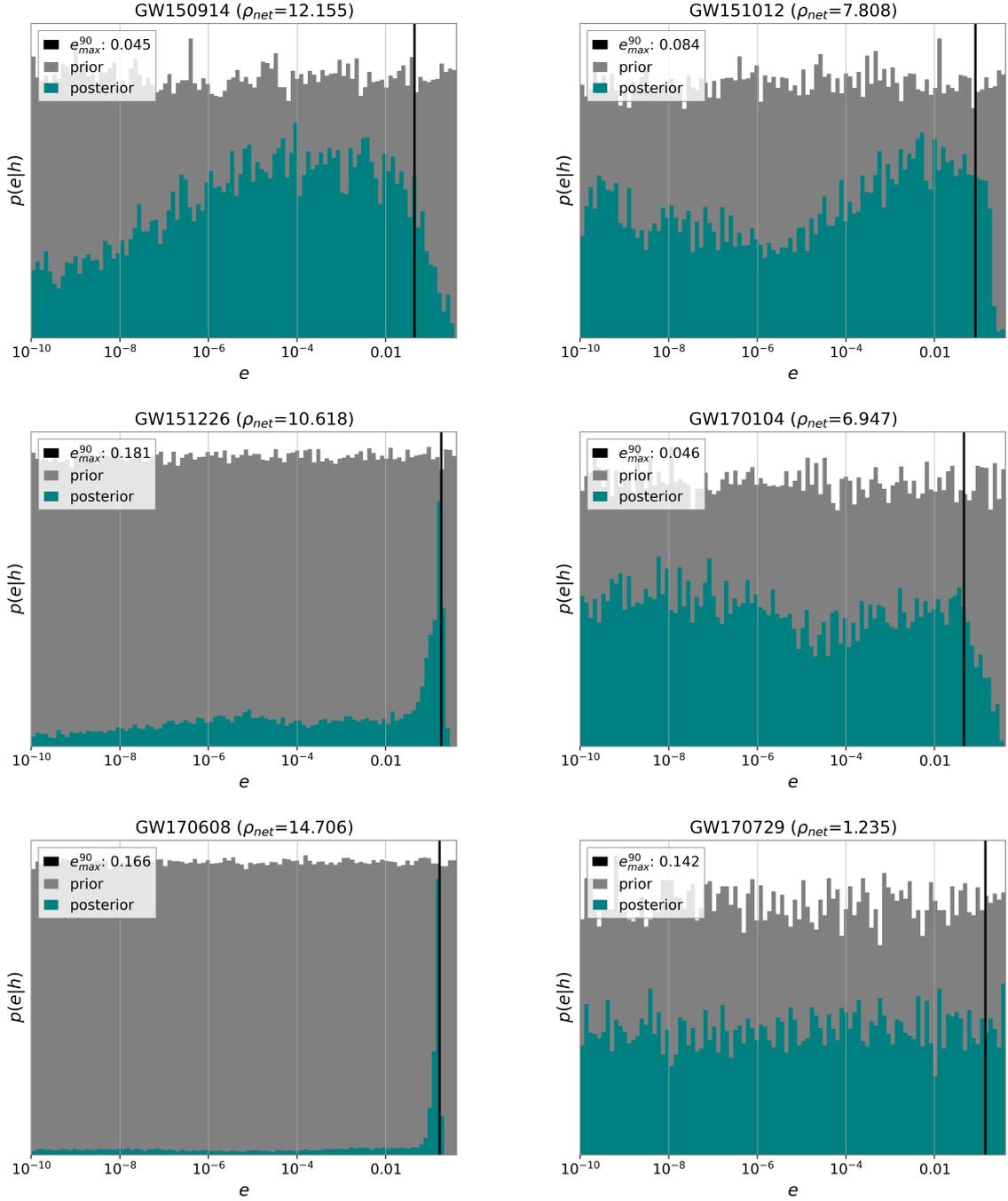

**Figure 7.** The posterior distribution of eccentricity for GW150914, GW151012, GW151226, GW170104, GW170608, GW170729 in `GWTC-1`. The green histogram represents the posterior distribution of eccentricity. The black line represents the upper limit of the 90% credible interval. The gray histogram is drawn from the prior distribution of eccentricity. In the title, $\rho_{\rm net}$ is the median value of the network matched filtering signal-to-noise ratio calculated with `EccentricFD`.

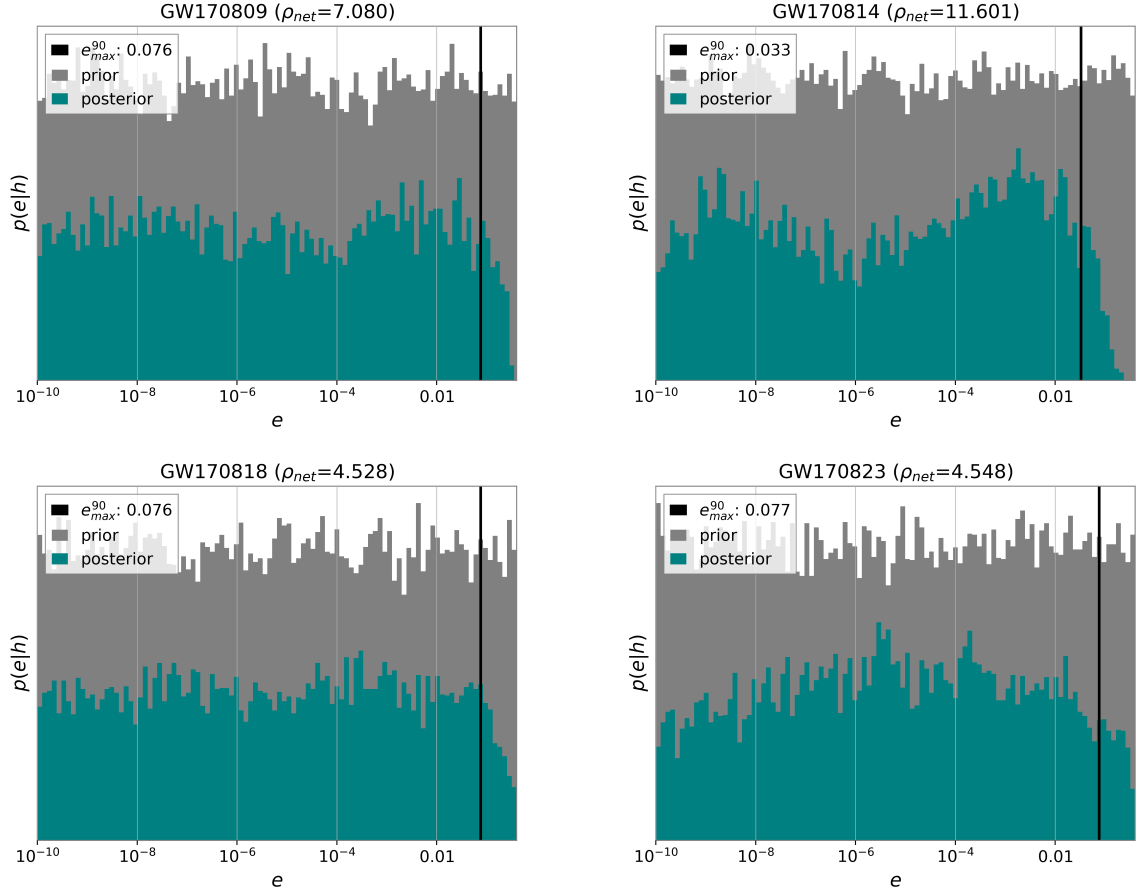

**Figure 8.** The posterior distribution of eccentricity for GW170809, GW170814, GW170818, GW170823 in `GWTC-1`. The plot convention is the same as that of the Figure 7.

**Table 5.** The prior distribution of `IMRPhenomPv2` simulated signal.

| parameter | unit | prior | range | boundary |
|---|---|---|---|---|
| $\mathcal{M}$ | $M_\odot$ | Uniform | [2, 60] | reflective |
| $q$ | 1 | Uniform | [0.125, 1] | reflective |
| $D_L$ | Mpc | UniformSourceFrame | [100, 5000] | reflective |
| $\phi$ | rad. | Uniform | $[0, 2\pi]$ | periodic |
| $\theta$ | rad. | Cosine | $[-\pi/2, \pi/2]$ | reflective |
| $\theta_{JN}$ | rad. | Sine | $[0, \pi]$ | reflective |
| $\psi$ | rad. | Uniform | $[0, \pi]$ | periodic |
| $\phi_c$ | rad. | Uniform | $[0, 2\pi]$ | periodic |
| $t_c$ | s | Uniform | $[t-0.1, t+0.1]$ | None |
| $a_1$ | 1 | Uniform | [0, 0.8] | reflective |
| $a_2$ | 1 | Uniform | [0, 0.8] | reflective |
| $\theta_1$ | rad. | Sine | $[0, \pi]$ | reflective |
| $\theta_2$ | rad. | Sine | $[0, \pi]$ | reflective |
| $\phi_{12}$ | rad. | Uniform | $[0, 2\pi]$ | periodic |
| $\phi_{JL}$ | rad. | Uniform | $[0, 2\pi]$ | periodic |

Combined with LIGO's $\chi_{\rm eff}$ measurement results on `GWTC-1` (Abbott et al. 2019b), the median value of $\chi_{\rm eff}$ of GW151226 and GW170729 is greater than 0.1, which cannot be ignored. According to the results of simulated signal, the eccentricity measurement results of these two BBH events are not reliable. Besides, the signal-to-noise ratio of GW170729's inspiral part is too low, the posterior of its eccentricity is almost the same as the prior. The median value of $\chi_{\rm eff}$ of GW170608 is only 0.03, which is lower than 0.1, but it can be seen from the Figure 9, GW170608 is an outlier, its large eccentricity may not be caused by neglecting $\chi_{\rm eff}$. According to Huerta et al. (2018), eccentric BBHs can be misclassified as quasicircular spinning BBHs after 15 Hz. LIGO used 30 Hz (H1) and 20 Hz (L1) as $f_{\rm min}$ on GW170608 (Abbott et al. 2017d), note that we measure the eccentricity of GW170608 at 10 Hz, so GW170608 is worthy of follow-up research. The distribution of the remaining 7 gravitational wave events is consistent with the simulated signals which $\chi_{\rm eff}$ less than 0.1, they are not outliers. Our results of these 7 GW events are in the same order of magnitude as the results of Romero-Shaw et al. (2019). Note that both we and Romero-Shaw et al. (2019) ignore the effect of $\chi_p$ on eccentricity. For now, the $\chi_p$ of `GWTC-1` is unconstrained (Abbott et al. 2019b). We will study the relationship between them in the near future.

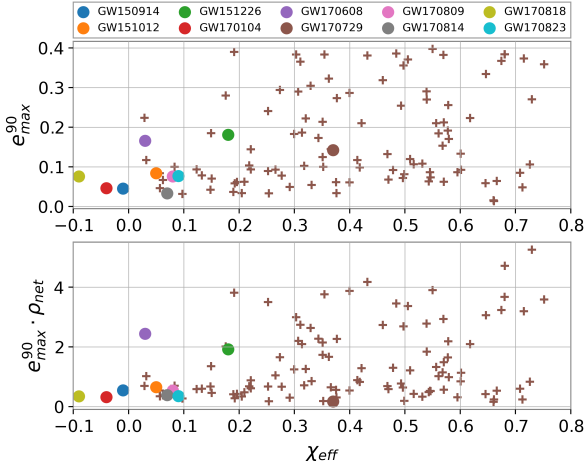

**Figure 9.** The scatter plot of the 100 simulated signals and 10 BBHs in GWTC-1. The brown marker '+' means the result of injected signal. '$\rho_{\text{net}}$' means the median network signal-to-noise ratio of parameter estimation results. Simulated signal's $\chi_{\text{eff}}$ is calculated from the injected parameters. GWTC-1's $\chi_{\text{eff}}$ is set by LIGO's results.

## 5 DISCUSSION

In this paper, we for the first time directly apply the stochastic sampling algorithm to measuring the eccentricity of the 10 BBH gravitational wave events in GWTC-1. Due to the slow waveform generation speed of SEOBNRE code, the inspiral-only waveform EccentricFD is adopted in the current work. The results show that the eccentricities of 7 BBHs of GWTC-1 (except for GW15226, GW170608 and GW170729) are very possibly less than 0.1 when the corresponding GWs enter the LIGO band. The relatively large eccentricity of GW151226 and GW170729 is probably due to ignoring the $\chi_{\text{eff}}$ and low signal-to-noise ratio, and GW170608 is worthy of follow-up research.

Due to the inspiral-only waveform, part of the GW signal is wasted. That is why the 90% credible interval of the eccentricity is relatively large. This indicates the importance of applying a complete eccentric BBH template to parameter estimation. Next step we plan to use SEOBNRE as the waveform template to do the measurement. In order to accelerate the waveform generation speed, there are some possible methods such as reduced order modeling (ROM) and reduced-order quadrature (ROQ) (Field et al. 2011, 2012; Canizares et al. 2015; Smith et al. 2016; Pürrer et al. 2017; Chua et al. 2019).

Previous work (Sun et al. 2015) indicates that using quasi-circular waveform templates may result in some eccentric BBH signals loss. Note that non-template search pipelines should have some sensitivity to eccentric BBH signals, but no significant eccentric BBH signal was found (Abbott et al. 2019e). Maybe when the complete inspiral-merger-ringdown spinning eccentric waveform template such as SEOBNRE is applied to matched-filtering search, we can find eccentric BBH events.

In the future, with the increasing sensitivity of gravitational wave detectors, the continuous accumulation of BBH events, and the improvement of eccentricity measurement methods, we will be able to distinguish the formation mechanisms of BBH by measuring the eccentricity of these BBHs.

## ACKNOWLEDGEMENTS

We thank Heng Yu for his great support, and we are thankful to Xiaolin Liu for many helpful discussions. This work was supported by the NSFC (No. 11622546, 11633001 and 11920101003) and by the Collaborative research program of the Institute for Cosmic Ray Research (ICRR), the University of Tokyo. Zhoujian Cao and Zong-Hong Zhu were supported by "the Interdiscipline Research Funds of Beijing Normal University" and the Strategic Priority Research Program of the Chinese Academy of Sciences, grant No. XDB23040100 and XDB23000000.

## REFERENCES

Aasi J., et al., 2013, Physical Review D, 88, 062001
Abbott B. P., et al., 2016a, Physical Review X, 6, 041015
Abbott B. P., et al., 2016b, Physical Review D, 93, 122003
Abbott B. P., et al., 2016c, Physical Review D, 93, 122004
Abbott B. P., et al., 2016d, Physical review letters, 116, 241102
Abbott B. P., et al., 2017a, Classical and Quantum Gravity, 34, 104002
Abbott B. P., et al., 2017b, Physical Review Letters, 119, 161101
Abbott B. P., et al., 2017c, The Astrophysical Journal Letters, 848, L13
Abbott B. P., et al., 2017d, The Astrophysical Journal Letters, 851, L35
Abbott R., et al., 2019a, arXiv preprint arXiv:1912.11716
Abbott B., et al., 2019b, Physical Review X, 9, 031040
Abbott B., et al., 2019c, Physical Review D, 100, 104036
Abbott B., et al., 2019d, The Astrophysical Journal Letters, 882, L24
Abbott B., et al., 2019e, The Astrophysical Journal, 883, 149
Abbott B., et al., 2020, arXiv preprint arXiv:2001.01761
Antelis J. M., Moreno C., 2019, General Relativity and Gravitation, 51, 61
Antonini F., Chatterjee S., Rodriguez C. L., Morscher M., Pattabiraman B., Kalogera V., Rasio F. A., 2016, ApJ, 816, 65
Ashton G., Khan S., 2019, arXiv preprint arXiv:1910.09138
Ashton G., et al., 2019, The Astrophysical Journal Supplement Series, 241, 27
Babak S., Taracchini A., Buonanno A., 2017, Physical Review D, 95, 024010
Bayes T., 1763, Philosophical transactions of the Royal Society of London, pp 370–418
Biwer C. M., Capano C. D., De S., Cabero M., Brown D. A., Nitz A. H., Raymond V., 2019, Publications of the Astronomical Society of the Pacific, 131, 024503
Bouffanais Y., Mapelli M., Gerosa D., Di Carlo U. N., Giacobbo N., Berti E., Baibhav V., 2019, The Astrophysical Journal, 886, 25
Buonanno A., Iyer B. R., Ochsner E., Pan Y., Sathyaprakash B. S., 2009, Physical Review D, 80, 084043
Canizares P., Field S. E., Gair J., Raymond V., Smith R., Tiglio M., 2015, Physical review letters, 114, 071104
Cao Z., Han W.-B., 2017, Physical Review D, 96, 044028
Casares J., Jonker P., 2014, Space Science Reviews, 183, 223
Celoria M., Oliveri R., Sesana A., Mapelli M., 2018, arXiv preprint arXiv:1807.11489
Chatziioannou K., Haster C.-J., Littenberg T. B., Farr W. M., Ghonge S., Millhouse M., Clark J. A., Cornish N., 2019, Physical Review D, 100, 104004
Christensen N., Meyer R., 2001, Physical Review D, 64, 022001
Chua A. J., Galley C. R., Vallisneri M., 2019, Physical review letters, 122, 211101
Cornish N. J., Crowder J., 2005, Physical Review D, 72, 043005
Corral-Santana J. M., Casares J., Munoz-Darias T., Bauer F. E., Martinez-Pais I. G., Russell D. M., 2016, Astronomy & Astrophysics, 587, A61
Cutler C., Flanagan E. E., 1994, Physical Review D, 49, 2658
De Mink S., Mandel I., 2016, Monthly Notices of the Royal Astronomical Society, 460, 3545
Dominik M., Belczynski K., Fryer C., Holz D. E., Berti E., Bulik T., Mandel I., O'Shaughnessy R., 2012, The Astrophysical Journal, 759, 52

Farr W. M., Stevenson S., Miller M. C., Mandel I., Farr B., Vecchio A., 2017, Nature, 548, 426

Farr B., Holz D. E., Farr W. M., 2018, The Astrophysical Journal Letters, 854, L9

Fernandez N., Profumo S., 2019, Journal of Cosmology and Astroparticle Physics, 2019, 022

Field S. E., Galley C. R., Herrmann F., Hesthaven J. S., Ochsner E., Tiglio M., 2011, Physical review letters, 106, 221102

Field S. E., Galley C. R., Ochsner E., 2012, Physical Review D, 86, 084046

Finn L. S., Chernoff D. F., 1993, Physical Review D, 47, 2198

Fragione G., Grishin E., Leigh N. W. C., Perets H. B., Perna R., 2019, MNRAS, 488, 47

Gayathri V., Bartos I., Haiman Z., Klimenko S., Kocsis B., Marka S., Yang Y., 2019, arXiv preprint arXiv:1911.11142

Ghosh A., et al., 2016, Physical Review D, 94, 021101

Ghosh A., et al., 2017, Classical and Quantum Gravity, 35, 014002

Gondán L., Kocsis B., Raffai P., Frei Z., 2018a, The Astrophysical Journal, 855, 34

Gondán L., Kocsis B., Raffai P., Frei Z., 2018b, ApJ, 860, 5

Hannam M., Schmidt P., Bohé A., Haegel L., Husa S., Ohme F., Pratten G., Pürrer M., 2014, Physical review letters, 113, 151101

Hastings W. K., 1970, Biometrika, 57, 97

Healy J., Lousto C. O., Zlochower Y., 2014, Physical Review D, 90, 104004

Hinder I., Vaishnav B., Herrmann F., Shoemaker D. M., Laguna P., 2008, Physical Review D, 77, 081502

Hinder I., Herrmann F., Laguna P., Shoemaker D., 2010, Physical Review D, 82, 024033

Hoang B.-M., Naoz S., Kocsis B., Rasio F. A., Dosopoulou F., 2018, ApJ, 856, 140

Huang Y., Middleton H., Ng K. K., Vitale S., Veitch J., 2018, Physical Review D, 98, 123021

Huerta E., Kumar P., McWilliams S. T., O'Shaughnessy R., Yunes N., 2014, Physical Review D, 90, 084016

Huerta E., et al., 2018, Physical Review D, 97, 024031

Husa S., Khan S., Hannam M., Pürrer M., Ohme F., Forteza X. J., Bohé A., 2016, Physical Review D, 93, 044006

Khan S., Husa S., Hannam M., Ohme F., Pürrer M., Forteza X. J., Bohé A., 2016, Physical Review D, 93, 044007

Klimenko S., Mitselmakher G., 2004, Classical and Quantum Gravity, 21, S1819

Klimenko S., Mohanty S., Rakhmanov M., Mitselmakher G., 2005, Physical Review D, 72, 122002

Klimenko S., Yakushin I., Mercer A., Mitselmakher G., 2008, Classical and Quantum Gravity, 25, 114029

Klimenko S., et al., 2016, Physical Review D, 93, 042004

Kowalska I., Bulik T., Belczynski K., Dominik M., Gondek-Rosinska D., 2011, Astronomy & Astrophysics, 527, A70

Kreidberg L., Bailyn C. D., Farr W. M., Kalogera V., 2012, The Astrophysical Journal, 757, 36

Kruckow M., Tauris T., Langer N., Szécsi D., Marchant P., Podsiadlowski P., 2016, Astronomy & Astrophysics, 596, A58

LIGO Scientific Collaboration 2018, LIGO Algorithm Library - LALSuite, free software (GPL), doi:10.7935/GT1W-FZ16

Liu X., Cao Z., Shao L., 2020, Physical Review D, 101, 044049

Livio M., Soker N., 1988, The Astrophysical Journal, 329, 764

Lower M. E., Thrane E., Lasky P. D., Smith R., 2018, Physical Review D, 98, 083028

Ma S., Cao Z., Lin C.-Y., Pan H.-P., Yo H.-J., 2017, Physical Review D, 96, 084046

Mandel I., De Mink S. E., 2016, Monthly Notices of the Royal Astronomical Society, 458, 2634

Messick C., et al., 2017, Physical Review D, 95, 042001

Metropolis N., Rosenbluth A. W., Rosenbluth M. N., Teller A. H., Teller E., 1953, The journal of chemical physics, 21, 1087

Moore B., Yunes N., 2019a, arXiv preprint arXiv:1910.01680

Moore B., Yunes N., 2019b, Classical and Quantum Gravity, 36, 185003

Nicholson D., Vecchio A., 1998, Physical Review D, 57, 4588

Nishizawa A., Sesana A., Berti E., Klein A., 2017, Monthly Notices of the Royal Astronomical Society, 465, 4375

Nitz A. H., Dal Canton T., Davis D., Reyes S., 2018, Physical Review D, 98, 024050

Nitz A. H., et al., 2019a, arXiv preprint arXiv:1910.05331

Nitz A. H., Lenon A., Brown D. A., 2019b, arXiv preprint arXiv:1912.05464

Nitz A. H., Capano C., Nielsen A. B., Reyes S., White R., Brown D. A., Krishnan B., 2019c, The Astrophysical Journal, 872, 195

O'Leary R. M., Rasio F. A., Fregeau J. M., Ivanova N., O'Shaughnessy R., 2006, ApJ, 637, 937

O'Leary R. M., Kocsis B., Loeb A., 2009a, MNRAS, 395, 2127

O'Leary R. M., Kocsis B., Loeb A., 2009b, Monthly Notices of the Royal Astronomical Society, 395, 2127

O'Leary R. M., Meiron Y., Kocsis B., 2016, ApJ, 824, L12

Pan Y., Buonanno A., Taracchini A., Kidder L. E., Mroué A. H., Pfeiffer H. P., Scheel M. A., Szilágyi B., 2014, Physical Review D, 89, 084006

Pan H.-P., Lin C.-Y., Cao Z., Yo H.-J., 2019, Physical Review D, 100, 124003

Payne E., Talbot C., Thrane E., 2019, Physical Review D, 100, 123017

Peters P. C., 1964, Physical Review, 136, B1224

Pürrer M., Smith R., Field S., Canizares P., Raymond V., Gair J., Hannam M., 2017, in The MG14 Meeting on General Relativity. pp 2015–2018

Raymond V., 2012, PhD thesis, Northwestern University

Rodriguez C. L., Farr B., Farr W. M., Mandel I., 2013, Physical Review D, 88, 084013

Romero-Shaw I. M., Lasky P. D., Thrane E., 2019, Monthly Notices of the Royal Astronomical Society, 490, 5210

Romero-Shaw I. M., Farrow N., Stevenson S., Thrane E., Zhu X.-J., 2020, arXiv e-prints, p. arXiv:2001.06492

Roulet J., Zaldarriaga M., 2019, Monthly Notices of the Royal Astronomical Society, 484, 4216

Sachdev S., et al., 2019, arXiv preprint arXiv:1901.08580

Samsing J., 2018, Phys. Rev. D, 97, 103014

Sidery T., et al., 2014, Physical Review D, 89, 084060

Sigurdsson S., Hernquist L., 1993, Nature, 364, 423

Skilling J., et al., 2006, Bayesian analysis, 1, 833

Smith R., Field S. E., Blackburn K., Haster C.-J., Pürrer M., Raymond V., Schmidt P., 2016, Physical Review D, 94, 044031

Speagle J. S., 2019, arXiv preprint arXiv:1904.02180

Stachie C., et al., 2020, arXiv preprint arXiv:2001.01462

Sun B., Cao Z., Wang Y., Yeh H.-C., 2015, Physical Review D, 92, 044034

Taam R. E., Sandquist E. L., 2000, Annual Review of Astronomy and Astrophysics, 38, 113

Tagawa H., Saitoh T. R., Kocsis B., 2018, Physical review letters, 120, 261101

Takátsy J., Bécsy B., Raffai P., 2019, MNRAS, 486, 570

Taracchini A., et al., 2014, Physical Review D, 89, 061502

Tetarenko B., Sivakoff G., Heinke C., Gladstone J., 2016, The Astrophysical Journal Supplement Series, 222, 15

The LIGO Scientific Collaboration the Virgo Collaboration 2020, arXiv e-prints, p. arXiv:2004.08342

Thrane E., Talbot C., 2019, Publications of the Astronomical Society of Australia, 36

Vallisneri M., 2008, Physical Review D, 77, 042001

Van Der Sluys M., Raymond V., Mandel I., Röver C., Christensen N., Kalogera V., Meyer R., Vecchio A., 2008, Classical and Quantum Gravity, 25, 184011

Veitch J., Vecchio A., 2010, Physical Review D, 81, 062003

Veitch J., et al., 2015, Physical Review D, 91, 042003

Venumadhav T., Zackay B., Roulet J., Dai L., Zaldarriaga M., 2019, arXiv preprint arXiv:1902.10341

Vitale S., Lynch R., Sturani R., Graff P., 2015, arXiv preprint arXiv:1503.04307

Vousden W., Farr W. M., Mandel I., 2016, Monthly Notices of the Royal Astronomical Society, 455, 1919

Wainstein L. A., Zubakov V., 1970, European Science Notes

Wen L., 2003, ApJ, 598, 419

Yunes N., Arun K., Berti E., Will C. M., 2009, Physical Review D, 80, 084001

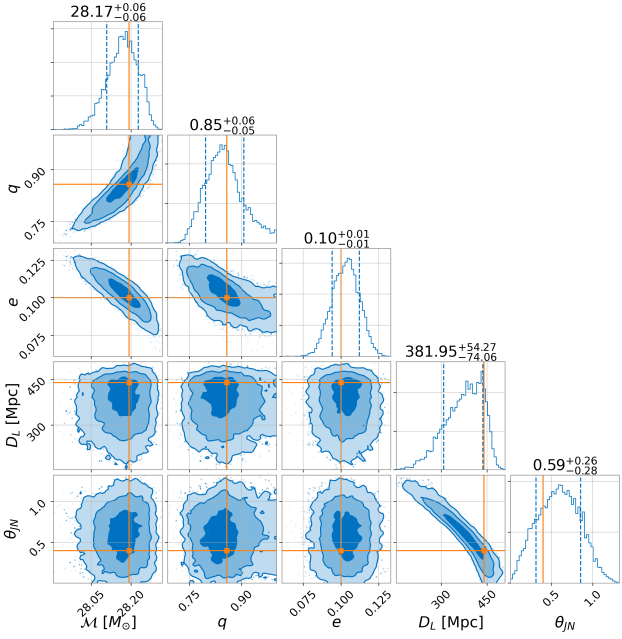

**Figure A1.** The posterior distributions of primary parameters for GW150914-like BBH with $e = 0.1$. The true values are given by the orange lines. The contours in the two-dimensional posteriors plot represent the 68%, 95%, and 99% credible intervals. The numbers listed at the top of each 1D marginal distribution plot are the median values of the posteriors and the boundary values of the 95% confidence intervals.

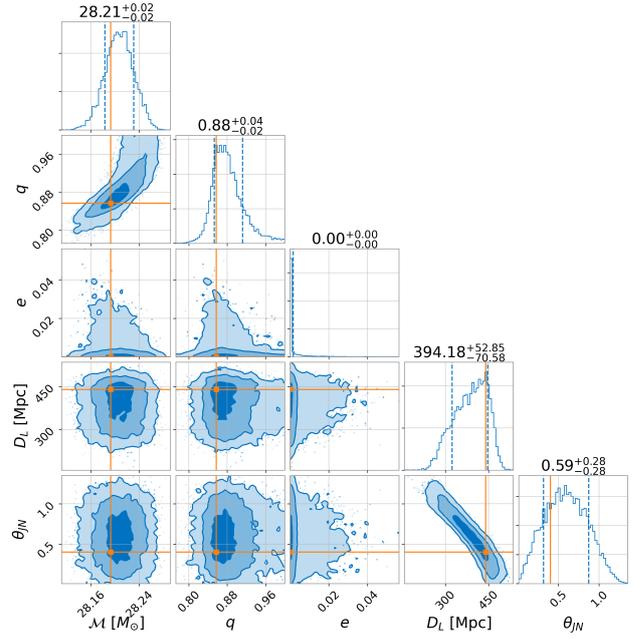

**Figure A2.** The posterior distributions of primary parameters for GW150914-like BBH with $e = 0.0$. The plot convention is the same as that of the Figure A1.

Zackay B., Dai L., Venumadhav T., Roulet J., Zaldarriaga M., 2019b, arXiv preprint arXiv:1910.09528
Zackay B., Venumadhav T., Dai L., Roulet J., Zaldarriaga M., 2019a, arXiv preprint arXiv:1902.10331

## APPENDIX A: THE POSTERIOR PROBABILITY OF THE SOURCE PARAMETERS OF TWO GW150914-LIKE BBHS

Here we provide the posterior distributions of primary parameters for the two GW150914-like BBHs illustrated in the Section 3.1.

This paper has been typeset from a TEX/LATEX file prepared by the author.